\documentclass[letterpaper, 10 pt, conference]{ieeeconf}  

\IEEEoverridecommandlockouts                              

\usepackage{graphicx}      
\usepackage{amssymb}
\usepackage{amstext}
\usepackage{amsmath}
\usepackage[ruled,vlined,algo2e]{algorithm2e}
\usepackage{booktabs}
\usepackage{ductri}
\usepackage{cancel}
\usepackage{float}

\title{\LARGE \bf
ANN-Based Adaptive NMPC for Uranium Extraction-Scrubbing Operation in Spent Nuclear Fuel Treatment Process*
}

\author{Duc-Tri Vo$^{1}$, Ionela Prodan$^{1}$, Laurent Lefèvre$^1$, Vincent Vanel$^2$, Sylvain Costenoble$^2$, Binh Dinh$^2$
\thanks{*The authors thank ORANO for partial financial support for the project}
\thanks{$^{1}$Univ. Grenoble Alpes, Grenoble INP, LCIS, 26000, Valence, France
        (e-mail: \{duc-tri.vo, ionela.prodan, laurent.lefevre\}@lcis.grenoble-inp.fr).}%
\thanks{$^{2}$CEA, DES, ISEC, DMRC, Univ Montpellier, Marcoule, France
        (e-mail: \{vincent.vanel, sylvain.costenoble, binh.dinh\}@cea.fr)}%
}

\begin{document}

\maketitle
\thispagestyle{empty}
\pagestyle{empty}

\begin{abstract}

This paper addresses the particularities in optimal control of the uranium extraction-scrubbing operation in the PUREX process. The control problem requires optimally stabilizing the system at a desired solvent saturation level, guaranteeing constraints, disturbance rejection, and adapting to set point variations. A qualified simulator named PAREX was developed by the French Alternative Energies and Atomic Energy Commission (CEA) to simulate liquid-liquid extraction operations in the PUREX process. However, since the mathematical model is complex and is described by a system of nonlinear, stiff, high-dimensional differential-algebraic equations (DAE), applying optimal control methods will lead to a large-scale nonlinear programming problem with a huge computational burden. The solution we propose in this work is to train a neural network to predict the process outputs using the measurement history. This neural network architecture, which employs the long short-term memory (LSTM), linear regression and logistic regression networks, allows reducing the number of state variables, thus reducing the complexity of the optimization problems in the control scheme. Furthermore, nonlinear model predictive control (NMPC) and moving horizon estimation (MHE) problems are developed and solved using the PSO (Particle Swarm Optimization) algorithm. Simulation results show that the proposed adaptive optimal control scheme satisfies the requirements of the control problem and provides promise for experimental testing.

\end{abstract}



\section{Introduction}\label{sec:introduction}
\subsection{Motivation}
The PUREX process, an acronym for "Plutonium, Uranium, Reduction, EXtraction," was developed to recover uranium and plutonium from spent nuclear fuels, which is composed of 95\% uranium, 1\% plutonium, and 4\% high radioactive toxic waste (the ultimate waste). This process offers a high-purity U-Pu recovery and recycling, reducing the ultimate waste volume and thus contributing to sustainable nuclear energy development. The overall control objective is quickly driving the process to achieve a desired solvent saturation level, guarantee constraints, handle the disturbances, and set point variations. 

PAREX is a simulation program developed by the French Alternative Energies and Atomic Energy Commission (CEA). It can simulate liquid-liquid extraction operations within the PUREX process. As reported in \cite{Bisson2016}, PAREX is currently used in the nuclear fuel reprocessing industry for process optimization, troubleshooting, and safety analysis. PAREX offers valuable insights into process dynamics and enables the applicability of model-based control approaches.

This work continues the studies of developing the adaptive Nonlinear Model Predictive Control (NMPC) for the uranium extraction-scrubbing operation in the PUREX process (\cite{vo2023}) and (\cite{Vo2023a}). We aim to exploit the benefits of the qualified PAREX simulator in the control scheme to satisfy the control objectives and constraints introduced above. However, it requires high-level security controls when developing an ANN replicate of PAREX since PAREX and its data are strictly protected. Therefore, in this first study, we propose a mathematical model that captures the main dynamics of the process, then use it to illustrate and study the developed control strategy in multiple simulations. Note that the proposed algorithm can be generalized to PAREX without any limitation.

In our previous studies (\cite{vo2023} and \cite{Vo2023a}), a high dimensional process model with 128 states was employed. However, note that from a practical viewpoint, only two state variables have critical roles in the control problem. Therefore, if we can reduce the number of variables in the process model, we can reduce the complexity of the control problem, which is the motivation of this paper.

Our main idea is to develop an artificial neural network (ANN) to predict the essential state variables based on available measurements. Then, the ANN is embedded as a predictor in the Nonlinear Model Predictive Controller (NMPC) scheme and as an estimator in the Moving Horizon Estimator (MHE) strategy. Furthermore, integrating NMPC and MHE allows us to have an adaptive control scheme in which any unmeasured disturbances can be estimated and updated to the controller. To solve the NMPC and MHE optimization problems, we use the enhanced Particle Swarm Optimization (PSO) developed in our previous work (\cite{Vo2023a}).

The Long Short-term Memory (LSTM) neural network, which was first proposed by \cite{Hochreiter}, is a common choice for time series prediction applications. Therefore, it represents a good candidate method for approximating system dynamics, allowing the application of model-based control techniques such as NMPC. The applicability of LSTM within NMPC was comprehensively discussed by \cite{JUNG2023106226}. Note that our proposed ANN architecture is based on the LSTM and linear and logistic regression networks. As will be discussed later in the paper, the ANN is designed based on the particularities of the control problem. 

\subsection{Contributions and Paper Organization}
This paper introduces an ANN-based adaptive control strategy tailored to the PUREX process's uranium extraction-scrubbing operation. The ANN architecture is developed based on LSTM, linear, and logistic regression networks. The ANN serves as a predictor and estimator in the control strategy, which NMPC and MHE solved using the enhenced PSO algorithm proposed in (\cite{Vo2023a}). Briefly, we:
\begin{itemize}    
    \item propose a dynamical model that captures the primary attributes of the uranium extraction-scrubbing operation in the PUREX process. In this paper, this model is used to develop and study the ANN-based adaptive MPC algorithm since the data sourced from PAREX is confidentially protected;

    \item develop an ANN that can predict the process outputs based on previous measurements. This ANN allows making predictions directly from the available measurements, which has significant advantages compared to the original state space approach since, in this process, not all the state values can be measured online;

    \item propose an adaptive control strategy that optimally drive the system to work at a desired solvent saturation level while guaranteeing constraints satisfaction (e.g., uranium concentration in the fission product, equipment limits) and adapting to parameters uncertainties (i.e., variation in the fresh solvent flow rate);
    
    \item study the efficiency of the proposed control strategy through simulations.
\end{itemize}

\textbf{Notations:} Vectors are denoted by bold lowercase letters and matrices are denoted by capital letters. $\bI,\, \mathbf{0}$ denotes identity and zero matrices of appropriate dimensions. $\bx_n$ denotes the $n^\text{th}$ element of $\bx$. $\bx_{n:m}$ denotes a vector slice from the $n^\text{th}$ to $m^\text{th}$ (included) elements of $\bx$. $\left\|\bx\right\|_\bQ := \bx^T\bQ\bx$. $\mathbb{N}_{m:n}:=\set{i \in \mathbb{N}| m\le i < n}$.

\begin{figure*}[ht]
    \centering
    \includegraphics[width=\linewidth]{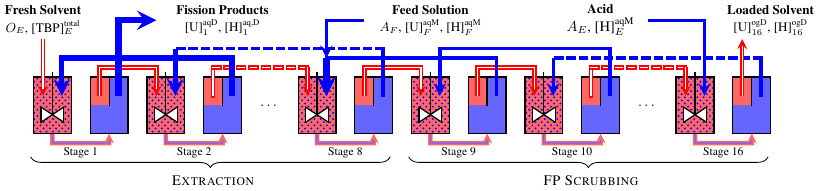}
    \caption{Uranium extraction-scrubbing operations using mixers-settlers (\cite{vo2023}).}
    \label{fig:system_detailed_EN}
\end{figure*}

\section{Uranium Extraction-Scrubbing Operation}
\subsection{Mathematical Model}\label{sec:mathematical_model}
Fig.~\ref{fig:system_detailed_EN} shows the uranium-scrubbing operation's principle consisting of 16 cascaded mixer settlers. In this work, we propose a model that captures the main dynamics of the qualified model in PAREX. This model is structured as a system of nonlinear, stiff, high-dimensional differential-algebraic equations (DAE). It adheres to mass balances and assumptions such as constant element densities, immiscible aqueous and organic phases, perfect mixing, and interfacial mass transfer derived from the double film theory. The main difference is that it uses a simplified distribution modulus, while PAREX's qualified modulus is derived from practical data. Notations for system parameters are detailed in Fig.~\ref{fig:system_detailed_EN} and Tab.~\ref{tab:system_params} while equation \eqref{eq:equilibrium} describes the primary extraction mechanism:

\begin{table}[H] 
    \centering
    \caption{System parameters.}
    \begin{tabular}{ll}
    \hline
    \textbf{Notation} & \textbf{Description} \\ \hline
    $A$, $O$ & Aqueous and organic flow rates.\\
    $V$, $W$ & Aqueous and organic volumes.\\
    $\KU$, $\KH$ & Equilibrium constants for $U$ and $H$.\\
    $k_U$, $k_H$ & Mass transfer coefficients for $U$, $H$.\\
    $^\text{a}$, $^\text{o}$ & Related to aqueous and organic phase.\\
    $^\text{M}$, $^\text{D}$ & Related to mixer and settler.\\
    $[\cdot]$ & Concentration.\\
    $_n$ & Related to stage $n$ of the process.\\
    $_{in}$ & Related to inputs to stage $n$. \\ 
    $_{i}$ & Related to concentrations at the interface.\\
    $_{*}$ & Related to concentrations at equilibrium.\\
    \hline
    \end{tabular}
    \label{tab:system_params}
\end{table}

\subsubsection{Interface mass transfer}
\begin{figure}[ht]
    \centering
    \includegraphics[width=0.7\linewidth]{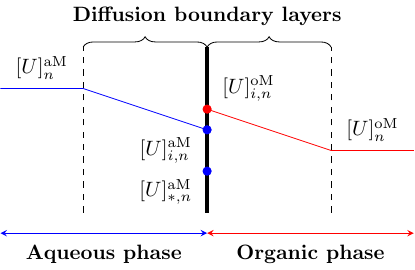}
    \caption{Profile of aqueous uranium concentration in the mixer, as derived with the two-film theory (\cite{Dinh2008b}).}
    \label{fig:two-film-theory}
\end{figure}
Applying the double film theory for uranium interfacial mass transfer, as illustrated in Fig.~\ref{fig:two-film-theory}, we have:
\begin{subequations}
\begin{gather}
    k_{U}^\text{aq} \left(\UaqM{n} - \UaqMi{n}\right)= k_{U}^\text{eq}\left(\UaqMi{n} - \UaqMis{n}\right),\\
    k_{U}^\text{aq} \left(\UaqM{n} - \UaqMi{n}\right)= k_{U}^\text{og}\left(\UogMi{n} - \UogM{n}\right).
\end{gather}\label{eq:double_film}
\end{subequations}
In this work, we assume that the transfer resistance can be neglectable, which can be expressed by using high mass transfer coefficients:
\begin{gather}
    k_{U}^\text{aq} = k_{U}^\text{eq} = k_{U}^\text{og} \gg 10^{-4} \text{ m/s}, \label{eq:mass_transfer_coeff}
\end{gather}
hence, \eqref{eq:double_film} can be rewritten as:
\begin{subequations}
\begin{align}
    \UaqMi{n} &= 0.5 \left(\UaqM{n} + \UaqMis{n}\right), \label{eq:uaqMi}\\
    \UogMi{n} &= 0.5\UaqM{n} + \UogM{n} - 0.5\UaqMis{n}. \label{eq:uogMi}
\end{align}\label{eq:u_relationship}
\end{subequations}
Similar relationships for nitric acid concentrations can be obtained by replacing U with H in equations \eqref{eq:double_film}-\eqref{eq:u_relationship}:
\begin{subequations}
\begin{align}
    \HaqMi{n} &= 0.5 \left(\HaqM{n} + \HaqMis{n}\right),\\
    \HogMi{n} &= 0.5\HaqM{n} + \HogM{n} - 0.5\HaqMis{n}. \label{eq:hogMi}
\end{align}\label{eq:h_relationship}
\end{subequations}

\subsubsection{Thermodynamic equilibrium}
The primary extraction mechanism is given by:
\begin{subequations}
\begin{align}
    \text{UO}_2^{2+} + 2\text{NO}_3^{-} + 2\text{TBP} &\overset{K_U}{\rightleftharpoons} \text{UO}_2\left(\text{NO}_3\right)_2\cdot \text{TBP},\label{eq:equi_eqs1}\\
    \text{H}^+ + \text{NO}_3^{-} + \text{TBP} &\overset{K_H}{\rightleftharpoons} \text{HNO}_3\cdot \text{TBP}.\label{eq:equi_eqs2}
\end{align}\label{eq:equilibrium}
\end{subequations}
At thermodynamic equilibrium condition, we have
\begin{subequations}
    \begin{align}
        \UogMi{n} &= \KU \UaqMis{n} \NOis{n}^2 \TBPfreei{n}^2,\\
        \HogMi{n} &= \KH \HaqMis{n} \NOis{n} \TBPfreei{n},
    \end{align}\label{eq:uogMi_hogMi}
\end{subequations}
where
\begin{subequations}
    \begin{align}
        \NOis{n} &= 2\UaqMis{n} + \HaqMis{n}, \label{eq:NO3}\\
        \TBPtotal{n} &= \TBPfreei{n} + 2 \UogMi{n} + \HogMi{n},\label{eq:TBPfree}
    \end{align}\label{eq:no3_tbp}
\end{subequations}
which leads to
\begin{subequations}
\begin{align}
    \TBPfreei{n} &= \dfrac{2c_n}{b_n + \sqrt{b_n^2 + 4a_nc_n}},
\end{align}\label{eq:tbp_free}
where
\begin{align}
    a_n &= 2\KU \UaqMis{n} \left(2\UaqMis{n}+\HaqMis{n}\right)^2,\\
    b_n &= 1 + \KH \HaqMis{n} \left(2\UaqMis{n}+\HaqMis{n}\right),\\
    c_n &= \TBPtotal{n}.
\end{align}
\end{subequations}
By substituting \eqref{eq:uogMi}, \eqref{eq:hogMi}, \eqref{eq:no3_tbp}, and \eqref{eq:tbp_free} to \eqref{eq:uogMi_hogMi}, and assume that $\TBPtotal{n} = \TBPtotal{E}$ for all stages, we obtain algebraic equations of uranium and acid concentrations, denoted by $g^U_n$, $g^H_n$, respectively:
\begin{subequations}
    \begin{align}
        g^U_n\left(\UaqM{n}, \UogM{n}, \HaqM{n}, \HogM{n}, \UaqMis{n}, \HaqMis{n}\right) &= 0,\\
        g^H_n\left(\UaqM{n}, \UogM{n}, \HaqM{n}, \HogM{n}, \UaqMis{n}, \HaqMis{n}\right) &= 0.
    \end{align}\label{eq:algebraic_equations}
\end{subequations}

\subsubsection{Flowrate and volume model}
\begin{figure}[H]
    \centering
    \includegraphics[width=0.8\linewidth]{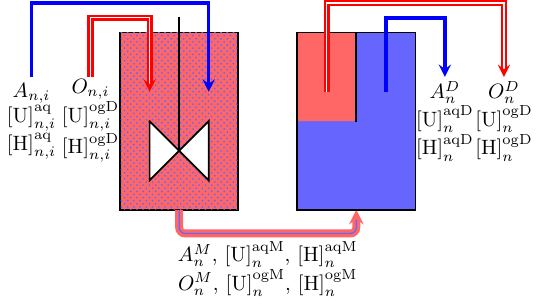}
    \caption{Mixer-settler model \cite{vo2023}.}
    \label{fig:SysDiagram_2}
\end{figure}

Consider the mixer-settler model showed in Fig.~\ref{fig:SysDiagram_2}, in the absence of monophasic reactions (only extraction phenomenon). The input flowrates and concentrations to the mixer of stage $n$ are:
\begin{subequations}
    \begin{align}
        A^M_{n,i} &= A^D_{n+1},& \quad O^M_{n,i} &= O^D_{n-1}, \\
        \UaqM{n,\text{in}} &= \UaqD{n+1},& \quad \UogM{n,\text{in}} &= \UogD{n-1}, \\
        \HaqM{n,\text{in}} &= \HaqD{n+1},& \quad \HogM{n,\text{in}} &= \HogD{n-1}.
    \end{align}
\end{subequations}
Note that at the stages $n\in \set{1, 8, 16}$, we have:
\begin{subequations}
\begin{align}
    O^M_{1,\text{in}}&=O_E,\; A^M_{8,i} = A^D_{9} + A_F, \;A^M_{16,\text{in}} = A_E,\\
    \UogM{1,\text{in}} &= \HogM{n,\text{in}},\; \UaqM{16,\text{in}} = 0,\; \HaqM{16,\text{in}} = \HaqE,\\
    \UaqM{8,\text{in}} &= \left(A_F \UaqF + A^D_9 \UaqD{9}\right)/(A_F + A^D_9),\\
    \HaqM{8,\text{in}} &= \left(A_F \UaqF + A^D_9 \HaqD{9}\right)/(A_F + A^D_9).
\end{align}
\end{subequations}

Applying total mass balances to the mixer tank, we have:
\begin{subequations}
    \begin{align}
        \dot{V}^M_n &= A^M_{n,i} - \dfrac{A^M_{n} + O^M_{n}}{\VM{n}+\WM{n}} \cdot \VM{n},\\
        \dot{W}^M_n &= O^M_{n,i} - \dfrac{A^M_{n} + O^M_{n}}{\VM{n}+\WM{n}} \cdot \WM{n}.
    \end{align}
\end{subequations}
The perfect mixing assumption mean $\dot{V}^M_n = \dot{W}^M_n = 0$, hence:
\begin{gather}
    A^M_{n,i} \WM{n} = O^M_{n,i} \VM{n}.
\end{gather}
In addition, assume that there is no leakages in the system and all mixer-settlers have the same volume, we have:
\begin{subequations}
\begin{gather}
    A_n^M = A_n^D, \quad O_n^M = O_n^D,\\
    \VM{1} + \WM{1} = \VM{2} + \WM{2} = \dots = \VM{16} + \WM{16}.
\end{gather}
\end{subequations}

\subsubsection{Mass balances of U and H}
Applying the mass balances to uranium in the aqueous and organic phases in the mixer tanks, we have:
\begin{subequations}
\begin{align}
    \dUaqM{n} &= \left(\An{n,i}[U]^\text{aq}_{n,i} - A^M_n \UaqM{n} - \Phi_n^U\right)/\VM{n},\\
    \dUogM{n} &= \left(\On{n,i}[U]^\text{og}_{n,i} - O^M_n \UogM{n} + \Phi_n^U\right)/\WM{n},\\
    \Phi_n^U &= \dfrac{6k_U \VM{n}}{d} \left(\UaqM{n}-\UaqMi{n}\right), \label{eq:phi_U}
\end{align}\label{eq:mass_balance_U_mixer}
where $\Phi_n^U$ denotes the uranium interfacial mass transfer term, and by substituting \eqref{eq:uaqMi} to \eqref{eq:phi_U}, we have
\begin{gather}
    \Phi_n^U = \dfrac{3k_U \VM{n}}{d} \left(\UaqM{n}-\UaqMis{n}\right). \label{eq:phi_U_2}
\end{gather}
\end{subequations}

Regarding the settler tanks, the mass balances equations for uranium are given as
\begin{subequations}
\begin{align}
    \dUaqD{n} &= \left(A^M_n \UaqM{n} - A^D_n\UaqD{n}\right) /\VD{n},\\
    \dUogD{n} &= \left(O^M_n \UogM{n} - O^D_n\UogD{n}\right) /\WD{n}.
\end{align}\label{eq:mass_balance_U_settler}
\end{subequations}

Mass balances equations for $H^+$ can be obtained by replacing $U$ by $H$ in \eqref{eq:mass_balance_U_mixer}-\eqref{eq:mass_balance_U_settler}.

\subsubsection{State space representation}
From \eqref{eq:algebraic_equations}, \eqref{eq:mass_balance_U_mixer}, and \eqref{eq:mass_balance_U_settler} , the process dynamical model can be represented as a system of differential-algebraic equations as follow:
\begin{subequations}
    \begin{align}
        \bdx &= \ff_c(\bx, \bx^\text{alg}, u, q),\\
        \mathbf{0} &= \bg(\bx, \bx^\text{alg}, u, q),
    \end{align}\label{eq:state_space}
where
\begin{itemize}
    \item $\bx \in \mathbb{R}^{128}$ is the vector of system states, i.e., uranium and acid concentrations:
    \begin{align*}
        \begin{array}{rcccc}
            \bx_{1:16} &= [\UaqM{1} & \UaqM{2} & \dots & \UaqM{16}]^T,\\
            \bx_{17:32} &= [\UogM{1} & \UogM{2} & \dots & \UogM{16}]^T,\\
            \bx_{33:48} &= [\UaqD{1} & \UaqD{2} & \dots & \UaqD{16}]^T,\\
            \bx_{49:64} &= [\UogD{1} & \UogD{2} & \dots & \UogD{16}]^T,\\
            \bx_{65:80} &= [\HaqM{1} & \UaqM{2} & \dots & \UaqM{16}]^T,\\
            \bx_{81:96} &= [\HogM{1} & \UaqM{2} & \dots & \UaqM{16}]^T,\\
            \bx_{97:112} &= [\HaqD{1} & \UaqM{2} & \dots & \UaqM{16}]^T,\\
            \bx_{113:128} &= [\HogD{1} & \UaqM{2} & \dots & \UaqM{16}]^T;
        \end{array}
    \end{align*}
    
    \item $\bx^\text{alg} \in \mathbb{R}^{32}$ is the vector of intermediate algebraic variables, i.e., uranium and acid concentrations:
    \begin{align*}
        \begin{matrix}
            \bx^\text{alg}_{1:16} &= [\UaqMis{1} & \UaqMis{2} & \dots & \UaqMis{16}]^T,\\
            \bx^\text{alg}_{17:32} &= [\HaqMis{1} & \HaqMis{2} & \dots & \HaqMis{16}]^T;
        \end{matrix}
    \end{align*}

    \item $u=A_F$ is the manipulated variable (the feed solution flowrate);

    \item $q=O_E$ is the disturbance variable (the fresh solvent flowrate);

    \item $\ff \in \mathbb{R}^{128}$ is the vector of mass balance equations;

    \item $\bg \in \mathbb{R}^{32}$ is the vector of algebraic equations;

    \item $y := \UaqD{9}=\bx_{41}$ is the controlled variable (the uranium concentration at the outlet of the scrubbing stage);

    \item $z := \UaqD{1} = \bx_{33}$ is the constrained variable (the uranium concentration in the fission products).
\end{itemize}
\end{subequations}
Several factors drive this choice of the controlled variable: i) not all the system states can be measured online; ii) these measurements usually have high cost; iii) this particular variable's value is compatible with our sensor's measurement range; and iv) analysis on the process dynamics show that it is a good indicator for the solvent saturation level. For convenience, we denote the discrete state space model as
\begin{subequations}
\begin{align}
    \bx (k+1) &= \ff_d \left(\bx(k), \bx^\text{alg}(k), u(k), q(k)\right),\\
    0 &= \bg \left(\bx(k), \bx^\text{alg}(k), u(k), q(k)\right),
\end{align}\label{eq:discrete_model}
\end{subequations}
where $\bx(k) := \bx(kT)$, and $T$ is the control sampling time of the overall system.

\subsection{Process Dynamics Analysis and Problem Statement}
\subsubsection{Solvent Saturation}
\begin{figure}[ht]
    \centering
    \includegraphics[width=0.9\linewidth]{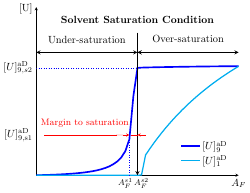}
    \caption{Steady state relationship of feed solution flow rates and uranium concentrations. Note that $\UaqD{9,s1} \approx 37.5\% \UaqD{9,s2}$.}
    \label{fig:SSC}
\end{figure}

\begin{figure}[ht]
    \centering
    \includegraphics[width=\linewidth]{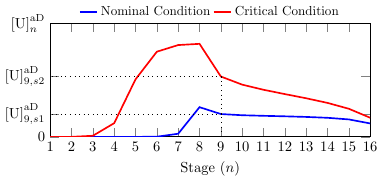}
    \caption{Profiles of aqueous uranium concentration in nominal and critical cases.}
    \label{fig:ssc_poc}
\end{figure}

An essential objective of the control system is to ensure that the system operates at a desired solvent saturation, which can be indicated by the aqueous uranium concentration at stage 9's settler. In general, a higher saturation level offers a higher decontamination towards fission products. However, there is a critical situation in which the solvent saturation is at maximum, which is indicated as $A_F^{s2}$ in Fig.~\ref{fig:SSC}. As solvent saturation increases, the profile aqueous uranium concentration in the settlers is shifted to the left, as illustrated in Fig.~\ref{fig:ssc_poc}.

The over-saturation situation is usually unwanted since it risks losing uranium to fission products, which reduces our extraction efficiency (since it is necessary to recover as much uranium as possible). Therefore, in nominal operation conditions, keeping the system operating in the under-saturated condition is preferable. This allows a safety margin towards the over-saturated region. Note that operating at a saturation condition is profitable in some cases: we can quickly flush out other elements. Therefore, the control system must also adapt to variations in the set point and guarantee that not much uranium leaks into the fission products.

\subsubsection{Problem Statement}\label{sec:problem_statement}
Considering the state space representation \eqref{eq:state_space}, the goal is to design an optimal control scheme that can regulate $y=\UaqD{9}$ to the desired set point $y_\text{set}=\UaqD{9,\text{set}}$ while accounting for unknown disturbances. Additionally, various constraints must be guaranteed during operation, such as aqueous uranium concentration in the extraction raffinates $\UaqD{1}$ \eqref{eq:cons_uaqD1}, overshoots (denoted by OS) \eqref{eq:cons_OS}, control input magnitudes $A_F$ \eqref{eq:cons_u}, and rates of control inputs \eqref{eq:cons_du}. The bounding values $A_F^\text{min}$, $A_F^\text{max}$, and $\Delta_{A_F}^\text{max}$ are defined based on the equipment limits, the flow sheet. In addition, since the constraints \eqref{eq:cons_uaqD1}-\eqref{eq:cons_u} have vital impacts on the overall control performance, we consider them as hard constraints, and if necessary, we allow \eqref{eq:cons_du} to be relaxed to guarantee the others:
\begin{subequations}
    \begin{gather}
        \UaqD{1} \le \UaqD{1,\text{tol}},\label{eq:cons_uaqD1}\\
        \text{OS} = y/y_\text{set} - 1 \le \text{OS}^\text{max}, \label{eq:cons_OS}\\
        A_F^\text{min} \le A_F \le A_F^\text{max}\label{eq:cons_u},\\
        \left| A_F(k+1) - A_F(k) \right|\le \Delta_{A_F}^\text{max}. \label{eq:cons_du}
    \end{gather}\label{eq:constraints}
\end{subequations}
Furthermore, we accept a steady state error of about 5\% of the set-point, i.e.,
\begin{gather}
    e = \left| y - y_\text{set}\right| \le 5\% y_\text{set}.\label{eq:steady_tolerance}
\end{gather}

\section{ANN-based Adaptive NMPC} \label{sec:ann_adaptive_nmpc}
\begin{figure}[ht]
    \centering
    \includegraphics[width=\linewidth]{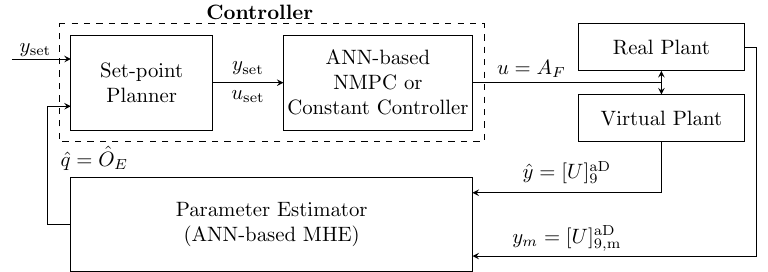}
    \caption{Proposed neural network architecture.}
    \label{fig:control_strategy}
\end{figure}

This section presents the development of our proposed ANN-based adaptive NMPC to stabilize the process output $y$ at a desired value $y_\text{set}$. The overall control system architecture is depicted in Fig.~\ref{fig:control_strategy}, where we first read measurements from the process and estimate the disturbance variable $\hq(k)$. This action allows the controller to have the best updates of the system parameters for further predictions. Within the controller block, we first compute the desired control input $u_\text{set}$ by solving the steady state condition:
\begin{subequations}
    \begin{align}
        \bdx &= \ff_c(\bx, \bx^\text{alg}, u, \hq),\\
        \mathbf{0} &= \bg(\bx, \bx^\text{alg}, u, \hq),\\
        y &= \bx_{41} = y_\text{set}.
    \end{align}
\end{subequations}
The optimal control input is then computed by the NMPC, which uses an ANN as the predictor. Note that if we keep the control input as constant, i.e., $u=u_\text{set}$, and the condition \eqref{eq:steady_tolerance} holds, we switch off the NMPC and apply directly $u_\text{set}$ to the system. This strategy helps to reduce computation costs, especially when we are near the steady state.

\subsection{ANN architecture} \label{sec:ann_architecture}
Regarding the control problem stated in Sec.~\ref{sec:problem_statement}, to implement the NMPC, two important outputs need to be predicted: $y:=\UaqD{9}$ and $z:=\UaqD{1}$, which are the aqueous uranium concentrations in the settlers of stage 9 and 1, respectively. In addition, since $z$ is involved only in the constraint \eqref{eq:cons_uaqD1}, it is not necessary to predict precisely the value of $z$. Instead, we can use a binary variable $\bar{z} 
\in \{0,1\}$, whose value is one if \eqref{eq:cons_uaqD1} holds and is zero otherwise. Furthermore, regarding the state space representation \eqref{eq:state_space}, since $y$ is the only online measurement that is available in the system, we aim to predict $y$ and $\bar{z}$ based on the history of $y$, $u := A_F$, and $p:=O_E$:
\begin{subequations}
    \begin{align}
        y(k+1) &= f_y\left(\btheta(k)\right),\\
        \bar{z}(k+1) &= f_{\bar{z}}\left(\btheta(k)\right),
    \end{align}
where the input vector $\btheta(k)$ is defined as:
    \begin{gather}
        \btheta(k) := \tmatright{\set{y(i)}_{i=k-N}^{k}\\ \set{u(i)}_{i=k-N}^{k}\\ \set{q(i)}_{i=k-N}^{k}}.
    \end{gather}\label{eq:ann_model}
\end{subequations}
The length of vector $\btheta(k)$ is $n_\theta = 3N$. Recall that $A_F$ and $O_E$ are the feed solution and fresh solvent flowrates, respectively.

\begin{figure}[ht]
    \centering
    \includegraphics[width=\linewidth]{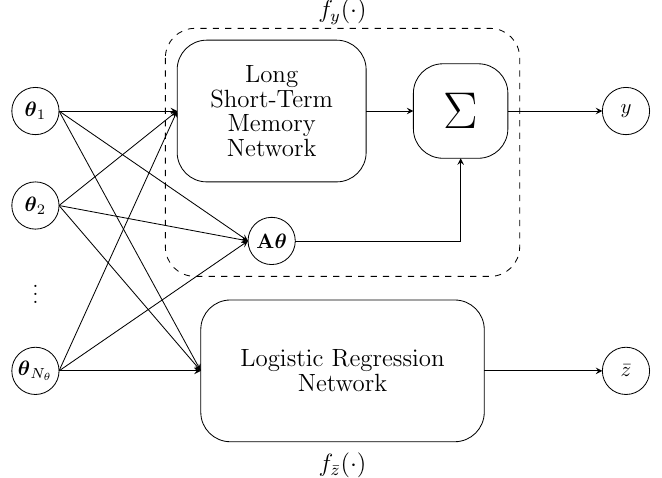}
    \caption{Proposed neural network architecture.}
    \label{fig:ann_architecture}
\end{figure}

The ANN architecture is depicted in Fig.~\ref{fig:ann_architecture}, in which $f_y$ is a linear regression model cascaded with an LSTM, and $f_{\bar{z}}$ is a binary classification model:
\begin{subequations}
\begin{align}
    f_y\left(\btheta(k)\right) &:= \bA \btheta(k) + f_\text{LSTM} \left(\btheta(k)\right),\label{eq:f1}\\
    f_{\bar{z}}\left(\btheta(k)\right) &:= f_\text{LG} \left(\btheta(k)\right).
\end{align}
\end{subequations}

The idea beyond $f_y$ is that we first predict $y$ by the linear regression model and then compensate for the linear prediction errors using the LSTM network. The training procedure for $f_y$ is described as follows:
\begin{enumerate}
    \item Run multiple simulations with the virtual plant, collect measurement histories, and construct the input-output sets $\mathcal{X},\, \mathcal{Y}_1$:
    \begin{gather}
        \mathcal{X} := \set{\btheta^{(i)}}_{i=0}^{N_\text{sam}}, \quad \mathcal{Y}_1 := \set{y^{(i)}}_{i=0}^{N_\text{sam}},
    \end{gather}
    where $N_\text{sam}$ is the number of data samples, $\btheta^{(i)}, y^{(i)}$ denote the $i^\text{th}$ input-output sample, respectively;
    \item Train the linear regression model with $\mathcal{X}$. In other words, find $\bA$ in \eqref{eq:f1};
    \item For each sample $i$, compute the linear regression error $e_1^{(i)}$ and construct the linear regression error set $\mathcal{E}_1$:
    \begin{gather}
        e_1^{(i)} := y^{(i)} - \bA \btheta^{(i)},\quad \mathcal{E}_1 := \set{e_1^{(i)}}_{i=0}^{N_\text{sam}};
    \end{gather}
    \item Train the LSTM model with $\mathcal{X}$ and $\mathcal{E}_1$.
\end{enumerate}
In addition, the logistic regression network $f_{\bar{z}}$ can be trained using the input set $\mathcal{X}$ and the output set $\mathcal{Y}_2$:
    \begin{gather}
        \mathcal{Y}_2 := \set{\bar{z}^{(i)}}_{i=0}^{N_\text{sam}}.
    \end{gather}

\subsection{Adaptive NMPC design} \label{sec:adaptive_nmpc}
To allow NMPC to have the ability to adapt to disturbances, which are unknown variations in the feed solution flow rate $\left(q = O_E\right)$, we develop a Moving Horizon Estimator to estimate this variable. This parameter estimator plays an essential role in the control strategy, i.e., it always allows the controller to be updated.

\subsubsection{Parameter Estimator} At time step $k$, denote $\hq(j|k),\, \hy(j|k)$ the estimated values of $q$ and $y$ at time step $j$. Suppose that we want to estimate $N_e$ latest values of $q$, which are denoted by $\set{\hq(i|k)}_{i=k-N_e}^{k-1}$, and assume that $N+N_e$ latest measurements $\set{y_m(i)}_{i=k-N-N_e+1}^k$ are available. Then, the best estimates of $\set{\hq(i|k)}_{i=k-N_e}^{k-1}$ can be obtained by solving the optimization problem:
\begin{subequations}
\begin{gather}
    \min_{\set{\hq(j|k)}_{j=k-N_e}^{k-1}} \sum_{j=k-N_e+1}^k \lambda^{k-j} \left(\hy(j|k) - y_m(j)\right)^2,
\end{gather}
such that $\forall j \in \mathbb{N}_{k-N_e:k-1}$:
\begin{align}
    \hy(j+1|k) &= f_y\left(\bhtheta(j|k)\right),\\
    \bhtheta(j|k) &:= \tmatright{
                    \set{\hy(i|k)}_{j-N}^{j}\\
                    \set{u(i)}_{j-N}^{j}\\
                    \set{\hq(i|k)}_{j-N}^{j}},\\
    \boldsymbol{\hat{\theta}}\left(k-N_e|k\right) &:=
    \tmatright{
        \set{y_m(i)}_{k-N_e-N}^{k-N_e}\\
        \set{u(i)}_{k-N_e-N}^{k-N_e}\\
        \set{\hq(i)}_{k-N_e-N}^{k-N_e}
        },
\end{align}\label{eq:optimization_problem_mhe}
where $\lambda \in (0,1)$ is the forgetting factor.
\end{subequations}

\textbf{Remarks:} The ANN-based MHE formulation developed above has $N_e$ decision variable. If the MHE is formulated based on the discrete mathematical model \eqref{eq:discrete_model}, the optimization problem will be:
    \begin{gather}
        \min_{\bhx\left(k-N_e\right),\,\set{\hq(j|k)}_{j=k-N_e}^{k-1}} \sum_{j=k-N_e+1}^k \lambda^{k-j} \left(\hy(j|k) - y_m(j)\right)^2.
    \end{gather}\label{eq:optimization_problem_mhe_128}
subject to process dynamics \eqref{eq:discrete_model}. In other words, we will have to estimate the initial condition $\bhx\left(k-N_e\right)$, which increases the number of decision variables by $128 \gg N_e$. Therefore, the optimization problem \eqref{eq:optimization_problem_mhe_128} requires a significant computation cost. An alternative approach was proposed in \cite{Vo2023a} where we keep track of the error history $\hy-y_m$ to decide which should be the initial condition $\bhx\left(k-N_e\right)$ for \eqref{eq:optimization_problem_mhe_128}. Therefore, the ANN allows us to make predictions directly based on the measurements and reduce the optimization problem's complexity, hence enabling practical implementation.

\subsubsection{NMPC Design} At time step $k$, denote $\set{\hu(i|k)}_{i=k}^{k+N_p-1}$ the sequence of predicted control inputs from time step $k$ to $k+N_p-1$. Here, $N_p$ denotes the prediction horizon length. In addition, assume that over the prediction horizon, the disturbance variable $q$ is constant, and its value equals $\hq(k-1) := \hq(k-1|k)$, computed by the parameter estimator. Next, by denoting $\hy(j|k)$ the predicted value of $y$ at time step $j$, the optimal predicted control inputs are obtained by solving the following optimization problem:

\begin{subequations}
    \begin{align}
        \min_{\set{u(i|k)}_{i=k}^{k+N_p-1}} &\sum_{j=k+1}^{k+N_p} w_p\left[\hu(j|k) - y_\text{set}\right]^2 + w_q\ty^2\left(N_p|k\right)\notag\\
        +& \sum_{j=k}^{k+N_p-1} w_r \left[\hu(j|k) - u_\text{set}\right]^2 \notag\\
        +& \sum_{j=k}^{k+N_p-1} w_s \left[\hu(j|k) - \hu(j-1|k)\right]^2
    \end{align}
subject to process dynamics, i.e. $\forall j \in \mathbb{N}_{k:k+N_p-1}$:
    \begin{gather}
        \hy(j+1|k) = f_y\left(\bhtheta(j|k)\right),\\
        \bar{z}(j+1|k) = f_{\bar{z}}\left(\bhtheta(j|k)\right),\\
        \bhtheta(j|k) := \tmatright{
                    \set{\hy(i|k)}_{k}^{j}\\
                    \set{\hu(i)}_{k}^{j}\\
                    \set{\hq(i|k)}_{k}^{j}};
    \end{gather}
constraints \eqref{eq:constraints}, i.e. $\forall j \in \mathbb{N}_{k:k+N_p-1}$:
    \begin{gather}
        \bar{z}(j|k) = 1,\\
        \hy(j|k) \le y_\text{set}\left(1+OS^\text{max} \right),\\
        A_F^\text{min} \le\hu(j|k) \le A_F^\text{max},\\
        -\Delta_{A_F}^\text{max}  \le\hu(j|k) - \hu(j-1|k) \le \Delta_{A_F}^\text{max};
    \end{gather}
and initial conditions:
    \begin{gather}
        \hu(k-1|k) := u(k-1),\\
        \hy(j|k) := y_m(j)\; \forall j \in \mathbb{N}_{k-N:k-1},\\
        \hq(i|k) := \hq(k-1)\; \forall j \in \mathbb{N}_{k:k+N_p-1}.        
    \end{gather}
\end{subequations}
Note that $w_p$, $w_q$, $w_r$, $w_s$ are weighting parameters chosen appropriately.

The optimization problems in MHE and NMPC can be solved using the extended PSO algorithm method developed in \cite{Vo2023a}. The main extension lies in handling the constraints \eqref{eq:constraints}. Briefly, whenever a candidate solution (a particle) breaks the constraints, we re-initialize it until all constraints are guaranteed.

\section{Case studies}
This section presents studies of the ANN-based Adaptive NMPC developed in Sec.~\ref{sec:ann_adaptive_nmpc}. The ANN was trained with \textit{Google Colab Pro+} resources and simulations were done on a laptop with \textit{Intel(R) Core(TM) i5-5200U CPU @ 2.20GHz} and 8GB RAM. For simulation of the real and virtual plant (cf. Fig.~\ref{fig:control_strategy}), we use the CasADi toolbox \cite{Andersson2019} with IDAS from SUNDIALS (\cite{hindmarsh2005sundials}, \cite{gardner2022sundials}) as the DAEs solver. In addition, the ANNs are trained using scikit-learn (\cite{scikit-learn}) and Tensorflow (\cite{tensorflow2015}).   The NMPC and MHE weighting coefficients are chosen heuristically. We will study 3 scenarios: i) the start-up period, ii) the switching to the critical case, and iii) the perturbed case. Simulation parameters are given in Tab.~\ref{tab:sim_params}.

\begin{table}[ht]
\caption{Simulation parameters.}
\begin{center}
\begin{tabular}{@{}llll@{}}
\toprule
\textbf{Parameter}         & \textbf{Value} & \textbf{Parameter} & \textbf{Value} \\ \midrule
$T$                        & 0.5 hours      & $N_e$              & 3 steps        \\
$\text{OS}^\text{max}$     & $20\%$         & $\lambda$          & 0.9            \\
$\varepsilon$              & $5\%$          & $N_p$              & 3 steps        \\
$N_\text{sam}$             & 4E6 samples    & $w_p=w_q$          & $1/y_\text{set}$              \\
Train-Validation-Test & 98\%-1\%-1\%   & $w_r=w_s$          & $1/u_\text{set}$              \\ \bottomrule
\end{tabular}\label{tab:sim_params}
\end{center}
\end{table}

\subsection{Data generation and ANN training}
For the simulations in this paper, we use the mathematical model developed in Sec.~\ref{sec:mathematical_model} to generate data and train the ANN. Firstly, we made several transient simulations with different values of $u$ and $q$. Then, we round all floating values to the six decimal places. For $N=2$, the training data has about four million records. Since the data set is big, we divide the train, validation, and test sets with a ratio of 98\%-1\%-1\%. The LSTM and logistic regression networks parameters and training results are given in Tab.~\ref{tab:ann_params}. 

\begin{table}[ht]
\caption{ANNs training information.}
\begin{center}
\begin{tabular}{@{}lll@{}}
\toprule
                        & LSTM                              & Logistic Regession                       \\ \midrule
Hidden layers           & 2                                 & 2                                        \\
Units per layer         & 10                                & 50                                       \\
Learning rate           & 1E-03                             & 1E-03                                    \\
Batch size              & 1024                              & 1024                                     \\
Epochs                  & 13                                & 5                                        \\
Final training loss     & 4.47E-4 (MAE) & 0.21 (Cross Entropy) \\
Final validation metric & 4.47E-4 (MAE) & 0.9 (Accuracy)       \\ \bottomrule
\end{tabular}\label{tab:ann_params}
\end{center}
\end{table}

\subsection{Start up period}
The system's initial condition is described as follows: We assume that uranium is only sent to the system since $t=0$. In other words, before $t=0$, only H and TBP are in the process, with system parameters equal to their nominal values. Furthermore, we assume that at $t=0$, the system is in a steady state (without uranium).

\begin{figure}[ht]
    \centering
    \includegraphics[width=\linewidth]{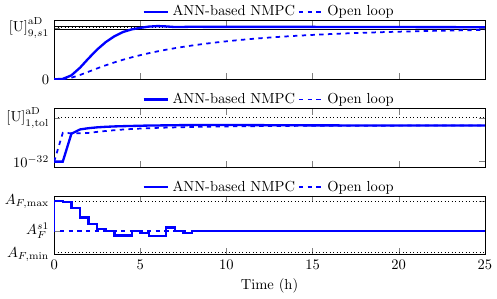}
    \caption{Simulation results of the start up period.}
    \label{fig:ann_nominal}
\end{figure}

Simulation results of the start-up period are shown in Fig.~\ref{fig:ann_nominal}. It can be seen that the proposed control strategy can effectively stabilize the system at the desired set point while guaranteeing all the constraints \eqref{eq:constraints}. Compared to the open loop controller, the controlled system significantly improves the settling time, up to 5 times faster. Furthermore, the switching condition \eqref{eq:steady_tolerance} is active since $t=8.5$h, which allows to turn off NMPC for a lower computation cost.

\subsection{Critical case}
\begin{figure}[ht]
    \centering
    \includegraphics[width=\linewidth]{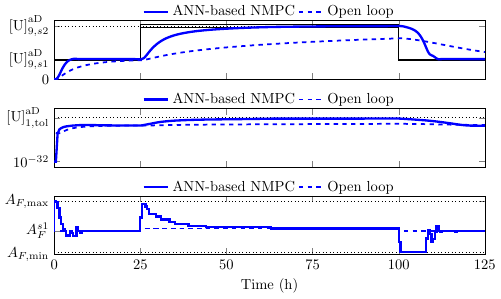}
    \caption{Simulation results when set point varies, i.e. switching to critical condition.}
    \label{fig:ann_sp_change}
\end{figure}

\begin{figure}[ht]
    \centering
    \includegraphics[width=\linewidth]{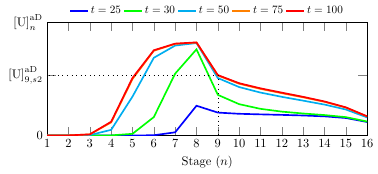}
    \caption{Profile of aqueous uranium concentration in settlers at certain time steps of the simulation in Fig.~\ref{fig:ann_sp_change}.}
    \label{fig:ann_sp_change_poc}
\end{figure}

To study the ability of the proposed control strategy to adapt to variations in the set point value, we continue the simulation in Fig.~\ref{fig:ann_nominal} and assume that $y_\text{set}$ increases to the critical value $\UaqD{9,s2}$ (cf. Fig.~\ref{fig:SSC}) at $t=25$h and decrease to its nominal value $\UaqD{9,s1}$ since $t=100$h. The simulation result of this case is depicted in Fig.~\ref{fig:ann_sp_change}. It can be seen that the controlled system can rapidly track the set point values while guaranteeing all the constraints. Furthermore, as expected, the uranium edge is shifted to the left when the solvent saturation increases, as depicted in Fig.~\ref{fig:ann_sp_change_poc}.

\subsection{Perturbed case}
\begin{figure}[ht]
    \centering
    \includegraphics[width=\linewidth]{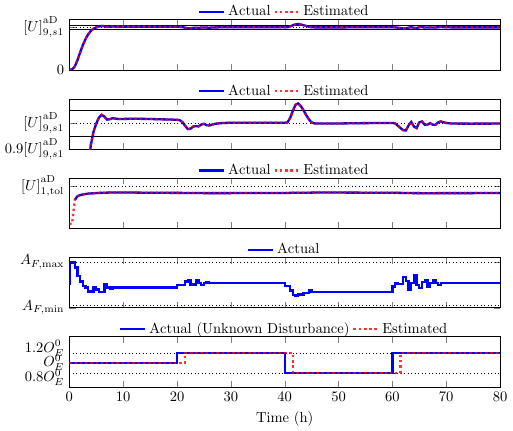}
    \caption{Simulation results of the perturbed case. Note that the first two subplots both show the behavior of $\UaqD{9}$ over time, but at different scales to give a better examination of this variable.}
    \label{fig:ann_perturbed}
\end{figure}

To study the disturbance rejection ability of the proposed control strategy, we analyze the case in which there are variations in the fresh solvent flow rate $O_E$. The simulation result for this case is shown in Fig.~\ref{fig:ann_perturbed}. It can be seen that the developed ANN-based MHE is capable of estimating the unknown disturbance, hence allowing the controller to be always updated with the process. Consequently, the controller can effectively adapt to unknown disturbances.

Additionally, we notice that there is always a delay of around 0.5h for the ANN-based MHE to make a successful estimation. This phenomenon can be explained by the slow dynamics of the process and the nature of our estimation algorithm. Regarding the MHE formulation \eqref{eq:optimization_problem_mhe}, the disturbance variable $q$ is estimated based on the output errors between the actual plant and its digital twin $\hy-y_m$. However, since the process dynamics are slow, it takes time for the effects of the disturbance to be reflected in the output, i.e., the error $\hy-y_m$ becomes sufficiently large. As a consequence, there are delays in parameter estimation. Furthermore, there are mismatches between the NMPC prediction model and the actual plant within these periods. Therefore, we noticed some violations at certain time steps: 20, 40, and 60h.

\section{Conclusions}
This paper presents an neural network based adaptive control strategy for the uranium extraction-scrubbing operation in the PUREX process. The artificial neural network (ANN), which includes the long-short term memory network, linear regression, and logistic regression networks, plays an essential role in the control strategy. Specifically, it is the predictor in the Nonlinear Model Predictive Control (NMPC) and the estimator in the Moving Horizon Estimation (MHE). The MHE and NMPC optimization problems are solved by our enhanced Partical Swarm Optimization (PSO) algorithm, proposed in \cite{Vo2023a}. ANN helps reduce the complexity of the MHE optimization problem, thus allowing online implementation ability. Multiple simulation case studies have shown that the developed control strategy is a candidate solution for our control problem: it can stabilize the system at a desired set-point while guaranteeing all the constraints, even in the critical condition or under unknown disturbances. Future developments include studies on other uncertainties handling, online learning, stability, robustness, and experimental implementation. Thorough comparisons with and without ANN will be carried out to assess the benefits of introducing ANN.

\addtolength{\textheight}{-12cm}   






\end{document}